\documentstyle[11pt,fleqn]{article}
\catcode`\@=11
\@addtoreset{equation}{section}
\def\theequation{\arabic{section}.\arabic{equation}}
\def\appendix{\renewcommand{\thesection}{\Alph{section}}\setcounter{section}{0}
              \renewcommand{\theequation}
            {\mbox{\Alph{section}.\arabic{equation}}}\setcounter{equation}{0}}
\def\maketitle{\thispagestyle{empty}\setcounter{page}0\newpage
                \renewcommand{\thefootnote}{\arabic{footnote}}
                  \setcounter{footnote}0}
\renewcommand{\thanks}[1]{\renewcommand{\thefootnote}{\fnsymbol{footnote}}
               \footnote{#1}\renewcommand{\thefootnote}{\arabic{footnote}}}
\newcommand{\preprint}[1]{\hfill{\sl preprint - #1}\par\bigskip\par\rm}
\renewcommand{\title}[1]{\begin{center}\Large\bf #1\end{center}\rm\par\bigskip}
\renewcommand{\author}[1]{\begin{center}\Large #1\end{center}}
\newcommand{\address}[1]{\begin{center}\large #1\end{center}}

\def\dinfn{\smallskip Dipartimento di Fisica, Universit\`a di Trento\\
                           and Istituto Nazionale di Fisica Nucleare,\\
                                   Gruppo Collegato di Trento, Italia}

\def\Idinfn{\address{\dinfn}}
\newcommand{\email}[1]{e-mail: \sl #1@science.unitn.it\rm}
\newcommand{\femail}[1]{\thanks{\email{#1}}}
\newcommand{\pacs}[1]{\smallskip\noindent{\sl PACS numbers:
                       \hspace{0.3cm}#1}\par\bigskip\rm}
\def\babs{\hrule\par\begin{description}\item{Abstract: }\it}
\def\eabs{\par\end{description}\hrule\par\medskip\rm}
\renewcommand{\date}[1]{\par\bigskip\par\sl\hfill #1\par\medskip\par\rm}
\newcommand{\ack}[1]{\par\section*{Acknowledgments} #1}
\newcommand{\s}[1]{\section{#1}}

\renewcommand{\vec}[1]{{\bf #1}}       %%%  vectors in bold
\newcommand{\ca}[1]{{\cal #1}}         %%%  calligraphic
\def\hs{\qquad}               %%%  horizontal space
\def\nn{\nonumber}            %%%  no number for eqnarray
\def\beq{\begin{eqnarray}}    %%%  begequation/eqnarray
\def\eeq{\end{eqnarray}}      %%%  endequation/eqnarray
\def\aq{\left[}               %%%  open [
\def\ag{\left\{}              %%%  open {
\def\cp{\right.}              %%%  close bracket
\def\cq{\right]}              %%%  close ]
\newtheorem{proposition}{Proposition}          %%% Proposition
\newtheorem{remark}{Remark}                    %%% Remark
\def\R{{\hbox{{\rm I}\kern-.2em\hbox{\rm R}}}}   %%% real numbers
\def\H{{\hbox{{\rm I}\kern-.2em\hbox{\rm H}}}}   %%% Hilbert space
\def\N{{\hbox{{\rm I}\kern-.2em\hbox{\rm N}}}}   %%% natural numbers
\def\C{{\ \hbox{{\rm I}\kern-.6em\hbox{\bf C}}}} %%% complex numbers
\def\Z{{\hbox{{\rm Z}\kern-.4em\hbox{\rm Z}}}}   %%% integers numbers
\def\ii{\infty}                                  %%% infinit
\newcommand{\fr}[2]{\mbox{$\frac{#1}{#2}$}}      %%% small fraction
\def\Tr{\mathop{\rm Tr}\nolimits}                  %%% Trace
\renewcommand{\Re}{\mathop{\rm Re}\nolimits}       %%% Real
\renewcommand{\Im}{\mathop{\rm Im}\nolimits}       %%% Imaginary
\def\lap{\Delta}                                   %%% Laplacian
\def\ga{\gamma}
\def\de{\delta}
\def\ep{\varepsilon}
\def\ze{\zeta}

\def\la{\lambda}

\def\om{\omega}

\def\Ga{\Gamma}

\def\La{\Lambda}

\def\citen#1{%
\edef\@tempa{\@ignspaftercomma,#1, \@end, }% ignore spaces in parameter list
\edef\@tempa{\expandafter\@ignendcommas\@tempa\@end}%
\if@filesw \immediate \write \@auxout {\string \citation {\@tempa}}\fi
\@tempcntb\m@ne \let\@h@ld\relax \let\@citea\@empty
\@for \@citeb:=\@tempa\do {\@cmpresscites}%
\@h@ld}
\def\@ignspaftercomma#1, {\ifx\@end#1\@empty\else
   #1,\expandafter\@ignspaftercomma\fi}
\def\@ignendcommas,#1,\@end{#1}
\def\@cmpresscites{%
 \expandafter\let \expandafter\@B@citeB \csname b@\@citeb \endcsname
 \ifx\@B@citeB\relax % undefined
    \@h@ld\@citea\@tempcntb\m@ne{\bf ?}%
    \@warning {Citation `\@citeb ' on page \thepage \space undefined}%
 \else%  defined
    \@tempcnta\@tempcntb \advance\@tempcnta\@ne
    \setbox\z@\hbox\bgroup % check if citation is a number:
    \ifnum\z@<0\@B@citeB \relax
       \egroup \@tempcntb\@B@citeB \relax
       \else \egroup \@tempcntb\m@ne \fi
    \ifnum\@tempcnta=\@tempcntb % Number follows previous--hold on to it
       \ifx\@h@ld\relax % first pair of successives
          \edef \@h@ld{\@citea\@B@citeB}%
       \else % compressible list of successives
          \edef\@h@ld{\hbox{--}\penalty\@highpenalty \@B@citeB}%
       \fi
    \else   %  non-successor--dump what's held and do this one
       \@h@ld \@citea \@B@citeB \let\@h@ld\relax
 \fi\fi%
 \let\@citea\@citepunct
}
\def\@citepunct{,\penalty\@highpenalty\hskip.13em plus.1em minus.1em}%
\def\@citex[#1]#2{\@cite{\citen{#2}}{#1}}%
\def\@cite#1#2{\leavevmode\unskip
  \ifnum\lastpenalty=\z@ \penalty\@highpenalty \fi % highpenalty before
  \ [{\multiply\@highpenalty 3 #1% % triple-highpenalties within list
      \if@tempswa,\penalty\@highpenalty\ #2\fi % and before note.
    }]\spacefactor\@m}
\begin{document}
%\tableofcontents       %%%%%%   index of section

\preprint{UTF 398,hep-th/9704035}

\title{Ray-Singer Torsion for a Hyperbolic 3-Manifold and Asymptotics
of Chern-Simons-Witten Invariant}

\author{Andrei A. Bytsenko\thanks{email: abyts@spin.hop.stu.neva.ru}}
\address{State Technical University, St. Petersburg 195251, Russia}
\author{Luciano Vanzo\femail{vanzo} and
Sergio Zerbini\femail{zerbini}}
\Idinfn

%\date{}

\babs
The Ray-Singer torsion  for a compact smooth hyperbolic 3-dimensional manifold
${\cal H}^3$ is expressed in terms of Selberg zeta-functions, making use of
the associated Selberg trace formulae. Applications to the evaluation of the
semiclassical asymptotics of the Witten's invariant for the Chern-Simons
theory with gauge group $SU(2)$ as well as to the sum over topologies in
3-dimensional quantum gravity are presented.

\eabs


\s{Introduction}

Recently, the topology of manifolds has been studied with the help of quantum
field theory methods. In this approach  the partition functions of quadratic functionals
play an important role. It has been shown that the Ray-Singer analytic torsion
(a topological invariant) \cite{ray71-7-145} can be obtained within quantum
field theory as the partition function of a certain quadratic functional
\cite{schw78-2-247,schw79-67-1}.

Furthermore, new invariants related to $3$-manifolds within the framework of
Chern-Simons gauge theory have been constructed in Ref. \cite{witt89-121-351}.
These invariants (well-defined topological invariants) have been specified in terms
of the axioms of topological quantum field theory \cite{moor89-123-77}.
The equivalent derivation of 3-manifold invariants has also been given
combinatorially in Ref. \cite{resh90-127-1,resh91-103-547}, where modular Hopf
algebras associated to quantum groups have been used.
A considerable interest in topological quantum field theory
has been stimulated by the introduction of the Witten's invariant. 
iThis invariant has been explicitly calculated
for a number of 3-manifolds and gauge groups
\cite{dijk90-129-393,kirb91-105-473,free91-141-79,jeff92-147-563,rama93-8-2285,
roza93u-99,roza94u-75,roza96-175-275}.

It has been observed that the semiclassical approximation can be associated
with the asymptotics $k\mapsto\infty$ of Witten's invariant $Z_W(k)$\,\,
(the level $k\in{\bf Z}$) of a $3$-manifold $M$ and gauge group $G$. Typically
this expression is a sum over partition functions of quadratic functionals.
For a number of $3$-manifolds and $G=SU(2)$ and  more general groups the
large-$k$ limit has been explicitly calculated in Refs. \cite{witt89-121-351,
free91-141-79,free91-66-1255,jeff92-147-563,roza93u-99,roza94u-75,
roza96-175-275,adam95u-95}.

The method considered in Ref. \cite{adam95u-95} for evaluating the
partition function (and the partition function appearing in the semiclassical
approximation) for a class of quadratic functionals is an extension and
refinement of the method proposed in \cite{schw78-2-247,schw79-67-1}. This
has led to the formulae for the partition function as a topological invariant
for a wider class of quadratic functionals. In addition, the Ray-Singer torsion
as a function of the cohomology \cite{ray73-23-167} has been obtained as the
partition function of a quadratic functional.

These methods have been applied in order to derive the usual Ray-Singer torsion
associated with a flat connection. When the cohomology of the connection is
non-vanishing the torsion is metric-dependent. However, in some cases, the
metric-dependence factors out in a simple way as a power of the volume of the
manifold to give a topological invariant \cite{adam95u-95}.

In this paper, we compute the Ray-Singer torsion for a compact smooth
3-dimensional hyperbolic manifold ${\cal H}^3\equiv H^3/\Ga$, \,$H^3$ being the
Lobachevsky space and $\Ga$ is a co-compact discrete group of isometries (for
more detail see \cite{byts96-266-1}). This result will be used for the
evaluation of the semiclassical asymptotics of the Witten's invariant related
to the manifold ${\cal H}^3$ and the gauge group $SU(2)$ as well as
for an expression of the one-loop 3-dimensional Euclidean partition function in the
case of negative cosmological constant.

The contents of the paper are the following.
In Sect.~2 we review the relevant information on the semiclassical
approximation in  the Chern-Simons theory, involving partition functions of
quadratic functionals. The analytic properties of zeta and eta functions
associated with elliptic operator acting on $p$-form are discussed in Sect. 3.
The partition function related to the elliptic resolvent and the trace formula
for the Laplacian on $p$-forms are investigated respectively in Sects. 4 and 5.
The new theoretical result of this paper, the explicit computation of
the Ray-Singer torsion and asymptotics of Witten's ${\cal H}^3$ invariant, is
presented in Sect. 6. We end with some conclusions in Sect. 7. Finally the
Appendices A and B contain a summary of the Plancherel measure, zeta
functions, and the Selberg trace formula for transverse vector fields.

\s{The Partition Function and Semiclassical Approximation}

In this Section, we briefly summarize the formalism we shall use in the
paper. To start with, we recall that the partition function of a quadratic
functional can be formally rewritten as follows
$$
Z(\beta)=\int_{\cal G}{\cal D}\omega e^{-\beta S(\omega)}\mbox{,}
\eqno{(2.1)}
$$
where $S(\omega)=F(\omega,\omega)$ is a real-valued quadratic functional on
the space ${\cal G}$ of sections $\omega$ (in a vectorbundle over a manifold
$M$) and $\beta$ is a complex-valued scaling parameter.

Let $\langle\cdot\,,\cdot\rangle$ b the inner product in ${\cal G}$ which
is determines an integration measure ${\cal D}\omega$. One can write
$S(\omega)=\langle\omega,T\omega\rangle$, where $T$ is a uniquely determined
self-adjoint operator.

Generally speaking, Eq. (2.1) is mathematically ill-definite and the partition
function is divergent, but  in our cases, since formally it  depends on
functional determinants of elliptic operators, it may be regularised via
zeta-function regularisation
\cite{hawk77-55-133}. If $S(\omega)$ is non-degenerate, $\mbox{ker}(T)=0$ then
one can obtain a finite expression for the partition function, which depends on
the choice of inner product $\langle\cdot\,,\cdot\rangle$ in ${\cal G}$.
If $S(\omega)$ is degenerate, the partition function diverges also due to the
divergent volume $V(\mbox{ker}(T))$, but it can be formally
expressed in terms of
$\zeta\equiv\mbox{dim}{\cal G}-\mbox{dim}(\mbox{ker}(T))$ and the (divergent)
volume $V(\mbox{ker}(T))$. A method for
evaluating the partition function in the degenerate case has been proposed in
\cite{schw78-2-247,schw79-67-1} and requires the functional $S(\omega)$ to
have an additional structure associated with it, namely a resolvent.

Let $({\cal G}_i,T_i)$ be a complex, i.e. a sequence of vector space
${\cal G}_p$ and linear operators $T_p$ acting from space ${\cal G}_p$ to the
space ${\cal G}_{p+1}$, $({\cal G}_{-1}={\cal G}_{N+1})$ and satisfying
$T_{p+1}T_p=0$ for all $p=0,1,...N$. Let us define the adjoint operators
$T_p^*:{\cal G}_{p+1}\mapsto{\cal G}_p$ by $\langle a,T_pb\rangle_{p+1}=
\langle T_p^*a,b\rangle_p$. A resolvent $R(S)$ of the functional $S$ (a chain
of linear maps) and the cohomology spaces, has the form
$$
0{\longmapsto}{\cal G}_N\stackrel{T_N}{\longmapsto}{\cal G}_{N-1}\stackrel
{T_{N-1}}{\longmapsto}...{\longmapsto}{\cal G}_1\stackrel{T_1}{\longmapsto}
ker(T){\longmapsto}0\mbox{,}
\eqno{(2.2)}
$$
$$
H^p(R(S))=\mbox{ker}(T_p)[\mbox{Im}(T_{p+1})]^{-1}\mbox{.}
\eqno{(2.3)}
$$
A generalisation of the Faddeev-Popov method given in Ref. \cite{schw79-67-1}
requires the cohomology of the resolvent to vanish. As a result the volume
$V(\mbox{ker}(T))$ can be evaluated from resolvent (2.2) in terms of the
divergent volums $V({\cal G}_p)$.

If the self-adjoint operators $\triangle_p:{\cal G}_p\mapsto{\cal G}_p$,
$$
\triangle_p=T_p^*T_p+T_{p-1}T_{p-1}^*\mbox{,}
\eqno{(2.4)}
$$
are elliptic differential operators, then the complex $({\cal G}_p,T_p)$ is an
elliptic complex. As anticipated, for the elliptic resolvent the determinants
associated with partition function can be regularised by standard
zeta-function regularization techniques. Thus a finite expression for the
partition function  will depend on the choice of resolvent (2.2) and inner
products $\langle\cdot\,,\cdot\rangle_p$ in the ${\cal G}_p$. For the elliptic
complex the spaces ${\cal G}_p$ are the spaces of smooth sections in
vectorbundles over the manifold $M$, the inner products in the ${\cal G}_p$
can be constructed from Hermitian structures in the bundles and a metric on
$M$. It has been shown \cite{schw79-67-1} that when the compact closed
manifold $M$ has odd dimension the partition function is invariant under
variation of the inner products in the ${\cal G}_p$ and under a certain
variation of the maps $T_p$ in the resolvent. In particular when
the definition of the functional $S$ and resolvent $R(S)$ does not require
choices of Hermitian structures or metric on $M$ the partition function is a
topological invariant. This results was generalised in Ref. \cite{adam95u-95}
to the case where the cohomology of the resolvent is non-vanishing.

We conclude this Sect. introducing the Witten's invariant, since
the method of evaluating the partition function of a quadratic functional can
be used in its the semiclassical approximation. The invariant is defined by
the partition function associated with  a Chern-Simons gauge
theory, i.e.
$$
Z_W(k)=\int {\cal D}Ae^{ikCS(A)}\mbox{,}\hspace{1.0cm} k\in Z\mbox{,}
\eqno{(2.5)}
$$
and
$$
CS(A)=\frac{1}{4\pi}\int_{M}\mbox{Tr}\left(A\wedge dA+\frac{2}{3}A
\wedge A\wedge A\right)\mbox{.}
\eqno{(2.6)}
$$
Since $CS(A)$ does not contain the metric on $M$, the quantity
$Z_W(k)$ is expected to be metric independent, namely to be a (well-defined)
topological invariant of $M$. Indeed, this fact has been proved in Refs.
 \cite{resh90-127-1,resh91-103-547}. The formal integration in (2.5) is over
the gauge fields $A$ in a trivial bundle, i.e. 1-forms on the 3-dimensional
manifold $M$ with values in Lie algebra ${\vec g}$ of the gauge group $G$.
In the formula (1.6) the group $G=SU(N)$ identified with its fundamental
representation, while the correct expression in the general case can be found
in Ref. \cite{free95-113-237}.

In the limit $k\mapsto\infty$, Eq. (2.5) is given by its semiclassical
approximation, involving only partition functions of quadratic functionals \cite{witt89-121-351}
$$
\sum_{[A_f]}e^{ikCS(A_f)}\int {\cal D}\omega e^{\frac{ik}{4\pi}\int_{{M}}
\mbox{Tr}(\omega\wedge d_{A_f}\omega)}\mbox{.}
\eqno{(2.7)}
$$
In above equation the sum is taken over representatives $A_f$ for each point
$[A_f]$ in the moduli-space of flat gauge fields on $M$. In addition the
$\omega$ are Lie-algebra-valued 1-forms and $d_{A_f}$ is the covariant derivative
determined by $A_f$, namely
$$
d_{A_f}\omega=d\omega+[A_f,\omega]\mbox{.}
\eqno{(2.8)}
$$

We shall use the method which enables the partition functions in  Eq. (2.7) to
be evaluated in complete generality (for more detail, see Ref.
\cite{adam95u-95}), namely the cohomology of $d_{A_f}$ is not required to
vanish. The general form of an action for each partition function of the
form (2.7) is
$$
S(\omega)=-\int_{M}\lambda_{\bf g}\mbox{Tr}(\omega\wedge d_{A_f}\omega),
\,\,\,\,\,\,\,\,
\beta=\frac{ik}{4\pi\lambda_{\bf g}}\mbox{,}
\eqno{(2.9)}
$$
where $\lambda_{\bf g}$ is an arbitrary parameter. The inner products in the
space $\Omega^q(M,{\vec g})$ of ${\vec g}$-valued q-forms naturally can be
chosen as $S(\omega)=\langle\omega,*d_{A_f(1)}\omega\rangle_{\lambda_{\bf g}}$,
\,\,$T\equiv*d_{A_f(1)}$. The canonical elliptic resolvent for quadratic
functional (2.9) is given by
$$
0\longmapsto\Omega^0(M,{\vec g})\stackrel{d_{A_f(0)}}{\longmapsto}\mbox{ker}
(d_{A_f(1)})\longmapsto 0\mbox{,}
\eqno{(2.10)}
$$
and the resolvent has cohomology spaces $H^0(R(S))=H^1(d_{A_f}),\,\,\,
H^1(R(S))=H^0(d_{A_f})$.

\s{Zeta and Eta Functions}

Let $\{\mu_j^{(p)}\}_{j=0}^{\infty}$ denote the non-zero eigenvalues (appearing
the same number of times as its multiplicity) of positive, selfadjoint Laplace
operators $\triangle_p$ and $\mu_j^{(p)}\leq\mu_{j+1}^p$ for all $j$.
The zeta function associated with operators $\triangle_p$,
$$
\zeta(s|\triangle_p)=\sum_j(\mu_j^{(p)})^{-s}\mbox{,}
\eqno{(3.1)}
$$
is well-defined analitic function for $\Re s >0$, $s\in{\vec C}$ and it can be
analytically continued to a meromorphic function on complex plane ${\vec C}$,
regular at $s=0$. The set $\{\mu_j^{(p)}\}_{j=0}^{\infty}$ is the union of the
non-zero eigenvalues of $T_p^*T_p$ and $T_{p-1}T_{p-1}^*$. It can be shown
that
$$
\{\mu_j^{(p)}\}=\{\lambda_l^{(p)}\}\cup\{\lambda_m^{(p-1)}\}\mbox{,}
\eqno{(3.2)}
$$
where $\{\lambda_l^{(p)}\}$ and $\{\lambda_m^{(p-1)}\}$ are the non-zero
eigenvalues of $T_p^*T_p$ and $T_{p-1}^*T_{p-1}$ respectively. In accordance
with the Ref. \cite{adam95u-95}, the zeta functions $\zeta(s|T_p^*T_p)$
$p=0,1,...,N$ are well-defined and analytic for $\Re(s)>0$ and can be
analytically continued to meromorphic functions on ${\vec C}$, regular at
$s=0$ and satisfy the formula
$$
\zeta(s|\triangle_p)=\zeta(s|T_p^*T_p)+\zeta(s|T_{p-1}^*T_{p-1})\mbox{.}
\eqno{(3.3)}
$$
One can define the heat kernel of elliptic operator
$$
\mbox{Tr}\left(e^{-t\triangle_p}\right)=\frac{-1}{2\pi i}\mbox{Tr}\int_{\gamma}
e^{-zt}(z-\triangle_p)^{-1}dz\mbox{,}
\eqno{(3.4)}
$$
where $\gamma$ is an arc on complex plane ${\vec C}$. By the standard result
in operator theory there exists $\epsilon,\delta >0$ such that for
$0<t<\delta$ the heat kernel expansion holds
$$
\mbox{Tr}\left(e^{-t\triangle_p}\right)=\sum_{0\leq l\leq l_0} a_l
(\triangle_p)t^{-l}+ O(t^\epsilon)\mbox{.}
\eqno{(3.5)}
$$
Starting with the formula \cite{adam95u-95}
$$
\zeta(0|\triangle_p)=a_0(\triangle_p)-\mbox{dim}(\mbox{ker}(\triangle_p))
=a_0(\triangle_p)
-\mbox{dim}H^p(R(S))\mbox{,}
\eqno{(3.6)}
$$
one can shown that the zeta function $\zeta(s||T|)$ ($|T|=\sqrt{T^2}$ is
defined via spectral theory) is well-defined and analytic for $\Re(s)>0$ and
can be continued to a meromorphic function on ${\vec C}$, regular at $s=0$ and
satisfies the formula
$$
\zeta(0||T|)=\sum_{p=0}^N(-1)^p(a_0(\triangle_p)-\mbox{dim} H^p(R(S)))\mbox{.}
\eqno{(3.7)}
$$
Finally the eta function
$$
\eta(s|\triangle_p)=\sum_j \mbox{sign}(\mu_j^{(p)})|\mu_j^{(p)}|^{-s}\mbox{,}
\eqno{(3.8)}
$$
is well-defined and analytic function for $\Re s>0$ and it can be analytically
continued to a meromorphic function on ${\vec C}$, regular at $s=0$.

\s{Quadratic Functional with Elliptic Resolvent}

Let $M$ be a compact oriented Riemannian manifold without boundary, and $n=2m+1
=\mbox{dim}M$ is the dimension of the manifold. Let the quadratic functionals
be defined on the space ${\cal G}={\cal G}(M,\xi)$ of smooth sections in a
real Hermitian vectorbundle $\xi$ over $M$.

Let $\chi:\pi_1(M)\longmapsto O(V,\langle\cdot\,,\cdot\rangle_V)$ be a
representation of $\pi_1(M)$ on real vectorspace $V$.
The mapping $\chi$ determines (on a basis of standard construction in
differential geometry) a real flat vectorbundle $\xi$ over $M$ and a flat
connection map $D$ on the space $\Omega(M,\xi)$ of differential forms on
$M$ with values in $\xi$. One can say that $\chi$ determines the space of
smooth sections in the vectorbundle $\Lambda(TM)^{*}\otimes \xi$.

Let $D_q$ denote the restriction of $D$ to the space
$\Omega^q(M,\xi)$ of $q$-forms and
$$
H^q(D)=\mbox{ker}(D_q)/\Im(D_{q-1})\mbox{,}
\eqno{(4.1)}
$$
are the cohomology spaces. A canonical Hermitian structure of the bundle
$\chi$ which $D$ is compatible with associated to $\langle\cdot\,,\cdot
\rangle_V$. The above mentioned Hermitian structure determines for each $x\in M$ a
linear map $\langle\cdot\,,\cdot\rangle_x: \xi_x\otimes\xi_x\longmapsto{\bf R}$,
and the diagram for linear maps hold (see Ref. \cite{adam95u-95} for details)
$$
\left(\Lambda^p(T_xM)^{*}\otimes\xi_x\right)\otimes
\left(\Lambda^q(T_xM)^{*}\otimes\xi_x\right)
\stackrel{\wedge}{\longmapsto}\Lambda^{p+q}(T_xM)^{*}\otimes
(\xi_x\otimes\xi_x)
\stackrel{\langle\cdot\,,\cdot\rangle_x}{\longmapsto}\Lambda^{p+q}
(T_xM)^{*}\mbox{,}
\eqno{(4.2)}
$$
where the image of $\omega_x\otimes\tau_x$ under this map has been denote by
$\langle\omega_x \wedge\tau_x\rangle_x$.

For odd $m$ we define the quadratic functional $S_D$ on $\Omega^m(M,\xi)$
by
$$
S_D(\omega)=\int_M\langle\omega(x)\wedge(D_m\omega)(x)\rangle_x
\mbox{,}
\eqno{(4.3)}
$$
where $D_m$ is the restriction of $D$ to $\Omega^m(M,\xi)$. One can
construct from the metric on $M$ and Hermitian structure in $\xi$ a Hermitian
structure in $\Lambda(T_xM)^{*}\otimes\xi$ and the inner products $\langle
\cdot\,,\cdot\rangle_q$ in the space $\Omega^q(M,\xi)$. Thus
$$
S_D(\omega)=\langle\omega,T\omega\rangle_m\mbox{,}\hspace{1.0cm}
T=*D_m\mbox{,}
\eqno{(4.4)}
$$
where $(*)$ is the Hodge-star map. Remind that the map $T$ is formally
selfadjoint with the property $T^2=D_m^{*}D_m$. For the functional
(4.3) there is a canonical topological elliptic resolvent $R(S_D)$,
$$
0\stackrel{0}{\longmapsto}\Omega^0(M,\xi)\stackrel{D_0}{\longmapsto}...
\stackrel{D_{m-2}}{\longmapsto}\Omega^{m-1}(M,\xi)\stackrel{D_{m-1}}
{\longmapsto}\mbox{ker}(S_D)\stackrel{0}{\longmapsto}0\mbox{.}
\eqno{(4.5)}
$$
From Eqs. (2.2) and (4.5) it follows that for the resolvent $R(S_D)$
we have $N=m, {\cal G}_p=\Omega^{m-p}(M,\xi), T_p=D_{m-p}$ and
$H^p(R(S_D))=H^{m-p}(D)$. Note that $S\geq 0$ and therefore
$\mbox{ker}(S_D)\equiv \mbox{ker}(T)=\mbox{ker}(D_m)$.

Let us choose an inner product $\langle\cdot\,,\cdot\rangle_{H^p}$ in each
space $H^p(R(S_D))$. The partition function of $S_D$ with the
resolvent (4.5) can be written in the form (see Ref. \cite{adam95u-95})
$$
Z(\beta)\equiv Z(\beta;R(S_D),\langle\cdot\,,\cdot\rangle_H,\langle
\cdot\,,\cdot\rangle)
=\pi^{\zeta/2}e^{-\frac{i\pi}{4}((\frac{2\theta}{\pi}\mp)\zeta\pm\eta)}|
\beta|^{-\zeta/2}\tau(M,\chi,\langle\cdot\,,\cdot\rangle_H)^{1/2}\mbox{,}
\eqno{(4.6)}
$$
where for $\beta=i\lambda, \lambda\in {\vec R}$, we have $\theta=\pm\pi/2$.
The function $\zeta$ appearing in the partition function above can be expressed
in terms of the dimensions of the cohomology spaces of $D$.

First of all, if the dimension of $M$ is odd then for all $p=0,1,...,N$,
$a_0=0$ in the asymptotic expansion (3.5). Since $H^p(R(S_D))=H^{m-p}(D)$
(the Poincar{\`e} duality) for the resolvent (4.5) it follows from Eq. (3.7)
that for $N=m$,
$$
\zeta\equiv\zeta(0||T|)=-\sum_{p=0}^N (-1)^p\mbox{dim} H^p(R(S))=(-1)^{m+1}
\sum_{q=0}^m (-1)^q\mbox{dim} H^q(D)\mbox{.}
\eqno{(4.7)}
$$

The factor $\tau(M,\chi,\langle\cdot\,,\cdot\rangle_H)$ is independent of
the choice of metric $g$ on $M$ \cite{adam95u-95}. In fact this quantity
is associated with the Ray-Singer torsion \cite{ray71-7-145} of the representation
$\chi$ of $\pi_1(M)$ constructed using the metric $g$. Thus $\tau(M,\chi,
\langle\cdot\,,\cdot\rangle_H)$ is a version of the Ray-Singer torsion as a
function of the cohomology defined and shown to be metric-independant in Ref.
\cite{ray73-23-167}. If $H^0(D)\neq 0$ and $H^q(D)= 0$ for
$q=1,...,m,\,\, n=2m+1$ is the dimension of $M$, then the product
$$
\tau(M,\chi,\langle\cdot\,,\cdot\rangle_H)=\tilde{\tau}(M,\chi,g)\cdot V(M)
^{-\mbox{dim} H^0(D)}\mbox{,}
\eqno{(4.8)}
$$
is independent of the choice of metric $g$, i.e. the metric dependence
of the Ray-Singer torsion $\tilde{\tau}(M,\chi,g)$ factors out as
$V(M)^{-\mbox{dim}H^0(D)}$.

The dependence of $\eta=\eta(0|T_D)$ on the connection map $D$
can be expressed with the help of formulae for the index of the twisted
signature operator for a certain vectorbundle over $M\otimes[0,1]$
\cite{atiy75-77-43}. It can be shown that \cite{adam95u-95}
$$
\eta(s|B^{(l)})=2\eta(s|T_{D^{(l)}})\mbox{,}
\eqno{(4.9)}
$$
where the $B^{(l)}$ are elliptic selfadjoint maps on $\Omega(M,\xi)$ defined on
$q$-forms by
$$
B_q^{(l)}=(-i)^{\lambda(q)}\left(*D^{(l)}+(-1)^{q+1}D^{(l)}*\right)
\mbox{,}
\eqno{(4.10)}
$$
In this formula $\lambda(q)=(q+1)(q+2)+m+1$ and for the Hodge star-map we have
used $*\alpha\wedge\beta=\langle\alpha,\beta\rangle_{vol}$. From the Hodge
theory
$$
\mbox{dim}\mbox{ker}B^{(l)}=\sum_{q=0}^n\mbox{dim}H^q(D^{(l)})\mbox{.}
\eqno{(4.11)}
$$
The metric-dependence of $\eta$ enters through $L^j(TM)$ and
$\eta(0|T_{D^{(0)}})$, where $L^j(TM)$ is the $j'$th term in Hirzebruch
$L$-polynomial (see for detail Ref. \cite{atiy75-77-43}) and $D^{(0)}$ is an
arbitrary flat connection map on $\Omega(M,\xi)$. For $n=3$ the only
contribution of the $L$-polynomial is $L_0=1$ and the metric-dependance of
$\eta$ is determined alone by $\eta(0|T_{D^{(0)}})$.

\s{Laplacian on Forms and Trace Formula}

In the application we shall consider a smooth compact 3-dimensional hyperbolic
manifold. The gauge group to be $G=SU(N)$ (its Lie algebra are identified with
their fundamental representations). Let us define an inner product in the Lie
algebra $\vec{g}$ by $\langle a,b\rangle_{\vec{g}}=-\lambda_{\vec{g}}\mbox{Tr}
(ab)$ with scaling parameter $\lambda_{\vec{g}}>0$. In the semiclassical
approximation, the partition functions are the partition functions of
functionals of the form $i\lambda S_{A_f}$, with $S_{A_f}=S_D $ given by Eq.
(2.9), $\xi=M\times {\vec{g}}$ with Hermitian structure determined by
$\langle\cdot\,,\cdot\rangle_{\vec{g}}$ and $D=d_{A_f}=d+ad(A_f)$ (where
$A_f$ is a flat gauge field and $ad:{\vec{g}}\mapsto\mbox{End}({\vec{g}})$ is
the adjoint representation, so the Eq. (2.8) holds).

In the following, we recall some results in the Hodge theory. Let $\de$ be the
Hodge operator acting on $p$-forms defined on a smooth 3-dimensional manifold
and $\lap=d\de+\de d$ is the Laplace operator. The following facts are well
known.

A transverse vector field is a co-closed one-form, i.e. $\de A=0$. The Hodge
decomposition of such a vector field is $A=\de\om+dJ+H$ for some
$2$-form $\om$, where $H$ is a harmonic and transverse vector field. Then the
condition $\de A=0$ would imply $\lap J=0$, which in turn is equivalent to
$J$ being a constant. Hence, modulo harmonic vector fields, each transverse
vector is actually co-exact.

If $\{\la^{(p)}\}_{l=0}^{\infty}$ are the eigenvalues of the operator $\de d$
restricted on $p$-forms and $\{\nu^{(p)}\}_{l=0}^{\infty}$ are the eigenvalues
of $d\de$ restricted on $p$-forms, then $\la^{(p)}_l=\nu^{(p+1)}_l$ with equal
multiplicity. It follows that if $A$ is transverse and $J$ is a closed
two-form such that $J=dA$ locally, then the eigenvalue problem
$\lap A\equiv\de dA=\la A$ gives all the eigenvalues of the problem
$\lap J\equiv d\de J=\la J$.

Let $\chi(\ga)$ be a character of $\Ga$, i.e. an homorphism $\chi(\ga):
\Ga\mapsto {\bf S}^1$. Then a twisted $p$-form, is one such that
$\ga^*A(x)=\chi(\ga)A(x)$ for any $\ga\in\Ga$, and we denote by $b_p$ the
number of twisted harmonic $p$-forms (twisted Betti numbers), i.e. the number
of twisted zero modes.

In order to obtain a trace formula for the Laplace-type operator acting on
transverse vector fields (starting from a trace formula for general vector
fields) we need to isolate the contribution of the scalar longitudinal mode.
More exactly, let $\lap_1^\perp$ and $\lap_0$ be the
Laplacians acting on transverse vector and scalar fields respectively.
Then for $A=A^\perp+dJ$, $\de A^\perp=0$, we have $\lap A=\lap_1^\perp
A^\perp+d\lap_0J$ and the same decomposition holds for polynomials (or possibly
for a class of well behaved functions). One can choose a basis $\{A_l\}$ for the
space of 1-forms and a set $A_l=A_l^\perp+dJ_l+H_l$, where $A_l^\perp$ is
transverse and $H_l$ is harmonic. This is the orthogonal Hodge decomposition
as applied to 1-forms, in view of the fact noted above that transverse vectors
are actually co-exacts.

Denoting $\lap_1$ the operator acting on vector fields (1-form), we have
\begin{eqnarray}
\Tr F(\lap_1^\perp)&=&\sum_l(A_l,F(\lap)A_l)
=b_1F(0)+\sum_l(A_l^\perp,F(\lap_1^\perp)A_l^\perp)+\sum_l
(dJ_l,F(\lap)dJ_l) \nn \\
&=&\Tr F(\lap_1^\perp)+\sum_l(dJ_l,F(\lap)dJ_l)\mbox{,}
\label{laptrac}
\end{eqnarray}
where $F$ is a suitable function and the crossed terms were zero by
orthogonality. Then the eigenvalues of operator $\lap$ on exact 1-forms are
equal to the eigenvalues of $\lap_0$ acting on the scalars which are the
divergence of a vector fields, i.e the co-exact $0$-forms with equal
multiplicity. Hence the last term in Eq.~(\ref{laptrac}) is the trace of
$F(\lap_0)$ on co-exact scalars. From the Hodge decomposition it follows that
each scalar is of the form $\phi=\delta \om_1+h$, where $h$ is harmonic.
Working as we did in Eq.~(\ref{laptrac}) it follows easily that
$\Tr F(\lap_0)=\Tr F(\lap_0)|_{(\mbox{co-exact})}+b_0F(0)$, where the first
trace is over all scalars and all non-vanishing eigenvalues. Hence we get
\beq
\Tr F(\lap_1^\perp)+(b_1-b_0)F(0)=\Tr F(\lap_1)-\Tr F(\lap_0)\mbox{.}
\label{trantrac}
\eeq
Thus $\Tr F(\lap_1)$ can be rewrite as the sum of $\Tr F(\lap_1^\perp)$ and
$\Tr F(\lap_0)$ plus the "number of zero modes" of $\lap_1^\perp$, namely the
sum associated with the orthogonal decomposition of the vector representation
of $SO(N)$ into irreducible summands\cite{frie86-84-523}.
Both traces in the right hand side of the Eq.~(\ref{trantrac}) are known in
terms of the geometric data of the manifold.

Let us generalize the above result to a trace formula for transverse p-form
defined on a smooth compact manifold. More generally our goal is to derive
the link between the trace of an arbitrary function $F(\lap^\perp)$, computed
by using constrained eigenfuntions of $\lap^\perp$, and the corresponding
unconstrained quantity. We shall consider completely antisymmetric tensors of
order $p$, that is $p$-forms $\om_p$. From the Hodge theory we have the
orthogonal decomposition
\beq
\om_p=\de\om_{p+1}+d\om_{p-1}+h_p
\:,\label{cc1}\eeq
where $h_p$ being a harmonic $p$-form, and the two equivalent eigenvalues
problems
\beq
\lap_p\om_p=\la\om_p\hs\Longleftrightarrow\hs\ag
\begin{array}{l}
\lap_{p+1}\,d\om_p=\la\,d\om_p\\
\lap_{p-1}\,\de\om_p=\la\,\de\om_p
\end{array}
\cp\:.\label{cc2}\eeq
This means that the spectra of the Hodge Laplacian acting on exact $p$-forms
and on co-exact $(p-1)$-forms are the same. The transverse part of the
antisymmetric tensor is represented by the co-exact $p$-form
$\om_p^{\perp}=\de\om_{p+1}$, which trivially satisfies $\de\om_p^{\perp}=0$,
and we denote by $\lap_p^\perp =\de d$ the restriction of the Laplacian on the
co-exact p-form.

Choosing a basis $\{\om_p^l\}$ of $p$-forms (eigenfunctions of the Laplacian)
we get
\beq
\sum_l \:\langle\om_p^l,F(\lap_p)\om_p^l\rangle=
\sum_l \:\langle\om_p^{l\perp},F(\lap_p)\om_p^{l\perp}\rangle
+\sum_l \:\langle d\om_{p-1}^l,F(\lap_p)d\om_{p-1}^l\rangle
+b_pF(0)
\:.\label{cc3}\eeq
Using the previous properties of $p$-forms and the Hodge Laplacian one can
obtain
\beq
&&\sum_l \:\langle d\om_{p-1}^l,F(\lap_p)d\om_{p-1}^l\rangle
=\sum_l \:\langle\de\,d\om_p^l,F(\lap_{p-1})\de\,d\om_p^l\rangle \nn \\
=&&\sum_l \:\langle\om_{p-1}^l,F(\lap_{p-1})\om_{p-1}^l\rangle
-\sum_l \:\langle d\om_{p-2}^l,F(\lap_{p-1})d\om_{p-2}^l\rangle
-b_{p-1}F(0)
\:.\label{cc4}\eeq
In this way we get
\beq
\Tr F(\lap_p^\perp) = \sum_{j=0}^{p} (-1)^j
\aq\Tr F(\lap_{p-j})\cq -\tilde{b}_{p} F(0)
\:,\label{TrF-pform}\eeq
where we have put
\beq
\tilde{b}_{p}=\sum_{j=0}^{p} (-1)^j b_{p-j}
\:.\label{betti}\eeq
Separating the $j=0$ term and using Eq.~(\ref{TrF-pform}), one has the trace
formula
\beq
\Tr F(\lap_p)=\Tr F(\lap_p^\perp)+\Tr 
F(\lap_{p-1}^\perp)+(\tilde{b}_{p}-\tilde{b}_{0}) F(0)
\:,\label{f}\eeq
which is the generalization of Eq.~(\ref{trantrac}).

Using the zeta function $\zeta(s|\lap_p)$ (see the Eqs. (3.1) and (3.3))
associated with operators $\lap_p$ and making the choice $F(x)=x^{-s}$, with
the convention $F(0)=0$, one can rewrite the Eq. (5.9) in the form
\beq
\ze(s|\lap_p)=\ze(s|\lap_p^\perp)+\ze(s|\lap_{p-1}^\perp)
\:.\label{f5}\eeq

Furthermore the definition of zeta-function regularized determinant, namely
\beq
\ln \det \lap_p=-\ze'(0|\lap_p)
\:,\label{bl}\eeq
gives
\beq
\det \lap_p=\det \lap_p^\perp \det \lap_{p-1}^\perp
\:.\label{f1}\eeq
Since $* \lap=\lap *$ the Hodge duality leads also to formula
\beq
\det \lap_p=\det \lap_{d-p}
\:.\label{f3}\eeq
As a consequence
\beq
\det \lap_p^\perp= \det \lap_{d-p-1}^\perp
\:,\label{f4}\eeq
and one can consider the spectral properties of the transverse Laplace
operator only.

\s{Ray-Singer Torsion and Asymptotics of Witten's ${\cal H}^3$ Invariant}

Recall the classification of all vector bundles over ${\bf S}^1$ for the
vector space ${\cal J}$ over fields ${\vec R}, {\vec C}$ and a body
${\vec H}$. For ${\cal J}={\vec R}$ there are trivial and 1-dimensional vector
bundles. Let $({\cal E},p,{\bf S}^1)$ be a bundle associated with
1-dimensional bundle. Then space ${\cal E}$ is a circle and the map $p:
{\cal E}\mapsto {\bf S}^1$ is a double covering. But for ${\cal J}={\vec C},
{\vec H}$ any vector bundle over base ${\bf S}^1$ is a trivial bundle.

The partition function (we consider vector line bundles over the manifold
${\bf S}^1\bigotimes {\cal H}^3$) is
$$
\left(2\pi\sqrt{\lambda_{{\bf g}}}\right)^{\zeta(A_f)}e^{-\frac{i\pi}{4}
\eta(A_f,{\bf g})}
k^{-\zeta(A_f)/2}\tau\left(M,A_f,<.,.>_{H(A_f)}\right)^{1/2}\mbox{,}
\eqno{(6.1)}
$$
where
$$
\zeta(A_f)=\mbox{dim}H^0(A_f)-\mbox{dim}H^1(A_f)\mbox{,}
\eqno{(6.2)}
$$
$$
\eta(A_f)=\eta(0|T_{A_f})\mbox{,}
\eqno{(6.3)}
$$
the quantity $\zeta(A_f)$ are formally given by $\zeta(0||T|)$ and
$\lambda_{\bf g}\equiv k(4\pi \lambda{\bf g})^{-1}$, $i\lambda\equiv \beta$.
Let the gauge group be $G=SU(2)$, and up to gauge equivalence $A_f=0$ is
the only flat gauge field on ${\cal H}^3$. Since $H^0(A_f)=su(2)$ and
$H^1(A_f)=0$, we get
$$
(2\pi)^3\lambda_{\bf g}^{3/2}k^{-3/2}\tau\left({\cal H}^3,A_f=0,<.,.>_{H^0(A_f=0)}
\right)^{1/2}V(SU(2))^{-1}\mbox{,}
\eqno{(6.4)}
$$
where we take the inner product in ${\bf g}=su(2)$ to be $<\bar{a},\bar{b}>=
-\lambda_{\bf g}\mbox{Tr}(ab)$,
$$
\tau\left({\cal H}^3,A_f=0,<.,.>_{H^0(A_f=0)}\right)^{1/2}=\tilde{\tau}
({\cal H}^3,\chi,g)^{3/2}V({\cal F})^{-3/2}\mbox{,}
\eqno{(6.5)}
$$
where $\tilde{\tau}({\cal H}^3,\chi,g)$ is the Ray-Singer torsion of
${\cal H}^3$, and $V({\cal F})$ is its volume.

The Ray-Singer torsion can be expressed in terms of the Selberg zeta functions
${\cal Z}_p(s)$ introduced in the Appendix B. To begin with, let us introduce
the Ruelle's zeta function in three dimension
$$
{\cal R}(s)=\prod_{p=0}^2{\cal Z}_p(p+s)^{(-1)^p}=\frac{{\cal Z}_0(s)
{\cal Z}_2(2+s)}{{\cal Z}_1(1+s)}\mbox{,}
\eqno{(6.6)}
$$
which can be extended to the entire complex plane as a meromorphic
function \cite{deit89-59-101}. The importance of Ruelle's zeta-function  lies
in the two theorems, first proved in \cite{frie86-84-523} for any
$N$-dimensional compact hyperbolic manifold. For the sake of completeness
we shall briefly discuss these aspects for the 3-dimensional case.

First let us introduce the Ray-Singer torsion of the $3$-manifold, associated
with a character $\chi$, by the formula
$$
\tilde{\tau}({\cal H}^3,\chi,g)=\frac{(\det\lap_1)^{1/2}(\det\lap_3)^{3/2}}
{\det\lap_2}\mbox{,}
\eqno{(6.7)}
$$
where $\lap_p$ is the laplacian restricted on $p$-forms and the determinants
are defined by means of zeta-function regularization. If the zero modes exist,
the determinants are to be defined by omitting the zero modes from the
Dirichelet series which are relevant to zeta functions.
By the Hodge duality analysis, presented in Sect. 5, we may also rewrite Eq.
(6.7) in the form
$$
\tilde{\tau}({\cal H}^3,\chi,g)=
\frac{(\det\lap_0)^{3/2}}{(\det\lap_1)^{1/2}}=\frac{(\det\lap_0)}{(\det\lap_1^
\perp)^{1/2}}\mbox{.}
\eqno{(6.8)}
$$
The Selberg trace formulae of Appendix B allow one to evaluate the analytic
continuations of the related zeta functions $\ze(s|\lap_p)$. If there are no
zero modes then \cite{byts96-266-1}
$$
\det \lap_0={\cal Z}_0(2)\exp\left(-\frac{V(\ca F)}{6\pi}\right)\mbox{.}
\eqno{(6.9)}
$$

For the transverse 1-form one may use again the correponding Selberg trace
formula. In addition, an intermediate regularization for the identity element
contribution should be performed, since the Plancherel measure in 3-dimension
case has no gap term (see for detail the Appendix A). Therefore let us
consider the Selberg trace formula (B.1) of the Appendix B and related to
the operator $\lap_1(m^2)\equiv\lap_1^\perp +m^2$. In the final formulae the
limit $m\to 0$ have to be taken. A straightforward computation leads to
$$
\ze(s|\lap_1(m^2))=\frac{V(\ca F)}{(4\pi)^{3/2}\Ga(s)}\left[(m)^{3-2s}\Ga(s-3/2)+
2(m)^{1-2s}\Ga(s-1/2)\right]+\frac{{\cal I}_1(s,m)}{\Ga(s)}\mbox{,}
\eqno{(6.10)}
$$
where
$$
{\cal I}_1(s,m)=\frac{1}{\Ga(1-s)}\int_0^\ii dt
(2tm+t^2)^{-s}\frac{{\cal Z}_1'}{{\cal Z}_1}(1+t+m)\mbox{.}
\eqno{(6.11)}
$$
Zeta-function regularization of the determinant yields
$$
\ln \det \lap_1(m^2)=-\ze'(0|\lap_1(m^2))=\frac{mV(\ca F)}{2 \pi}\aq
\frac{m^2}{3}-1 \cq+\ln {\cal Z}_1(1+m)\mbox{.}
\eqno{(6.12)}
$$
For the vanishing Betti number $\tilde{b_1}$ the limit $m \to 0$ gives
$$
\det \lap_1^\perp={\cal Z}_1(1)\mbox{.}
\eqno{(6.13)}
$$
As a consequence, we have the Ray-Singer torsion in the form
$$
\tilde{\tau}^2({\cal H}^3,\chi,g)=\frac{{\cal Z}_0(2)^2}{{\cal Z}_1(1)}
\exp\left(-\frac{V({\cal F})}{3\pi}\right)\mbox{.}
\eqno{(6.14)}
$$
In this form the dependence on the volume has been extracted. In
fact, if one computes the Ray-Singer torsion for $H^3$, one naively
obtains $\tilde{\tau}^2(H^3) \simeq
\exp\left(-\frac{V_3}{3\pi}\right) $, with $V_3$ very large, and the
result is zero in the limit $V_3 \mapsto \ii$.

It should be noted that we can rewrite the torsion as
$$
\tilde{\tau}^2({\cal H}^3,\chi,g)=\frac{{\cal Z}_0(0)^2}{{\cal Z}_1(1)}
\exp\left(\frac{V({\cal F})}{3\pi}\right)={\cal R}(0)\mbox{,}
\eqno{(6.15)}
$$
where the use of functional equation (\ref{zfuncsca}) has been made,
namely
$$
{\cal Z}_0(2)={\cal Z}_0(0)\exp\left(\frac{V(\ca F)}{3\pi}\right)\mbox{.}
\eqno{(6.16)}
$$
This result is a particular case of Fried's first theorem \cite{frie86-84-523}
in three dimensions: when the twisted Betti's numbers all vanish, one has
$\tilde{\tau}({\cal H}^3,\chi,g)=|{\cal R}(0)|^{1/2}$.

The second theorem (in three dimensions) states that in the presence of
non-vanishing Betti numbers, the leading term in the Laurent expansion of
${\cal R}(s)$ around $s=0$ is
$$
(-4)^{b_0}\tilde{\tau}^2({\cal H}^3,\chi,g)s^{4b_0-2b_1}\mbox{.}
\eqno{(6.17)}
$$
So the analytic torsion can still be identified as the leading term of
the Ruelle zeta-function at $s=0$. Then since
$$
\det \lap_0=\frac{1}{b_0!}{\cal Z}_0^{(b_0)}(2)\exp\left(-\frac{V({\cal F})}{6\pi}
\right)\mbox{,}
\eqno{(6.17)}
$$
and for $\tilde{b}_1=b_1-b_0 >0$,
$$
\det \lap_1^\perp=\frac{1}{\tilde{b}_1!} {\cal Z}_1^{(\tilde{b}_1)}(1)\mbox{,}
\eqno{(6.18)}
$$
one may also write
$$
\tilde{\tau}^2({\cal H}^3,\chi,g)=\frac{\tilde{b}_1!({\cal
Z}_0^{(b_0)}(2))^2}
{(\tilde{b}_0!)^2 {\cal Z}_1^{(\tilde{b}_1)}(1)}\exp\left(-\frac{V({\cal F})}
{3\pi}\right)\mbox{.}
\eqno{(6.19)}
$$

We conclude this Sect. with some remarks. It is an useful fact that in three
dimensions one has ${\cal Z}_2(s)={\cal Z}_0(s)$, essentially because
$\La^2{\bf C}^2$ is isomorphic to $\La^0{\bf C}^2$. In the second case
(non-vanishing Betti numbers) the functional equation seems useless. Besides
one may note that for non-trivial characters $b_0=0$ and possibly also $b_1=0$,
so one falls in previous case. For trivial character, on the other hand, one
has $b_0=1$ (for any closed manifold) and $b_1=0$ for an infinite
number of ${\cal H}^3$, so that ${\cal R}(s)$ has a zero at $s=0$ of
order $4$. However, it is known that there exist a class of compact
hyperbolic manifolds which admits arbitrarly large value of $b_1$, the
so called Haken class.

As a consequence, the asymptotics of the Witten's invariant for a
smooth compact hyperbolic 3-manifold has been expressed as a function of the
Selberg zeta-functions ${\cal Z}_p$.

\s{Conclusions}

In this paper the Ray-Singer torsion for a 3-dimensional compact hyperbolic
manifold has been evaluated as a function of the Selberg zeta-functions
and the volume of the fundamental domain. As first application,
we have derived explicit formulae for the Chern-Simons-Witten invariant
related to this manifold for arbitrary values of the level $k$
making use of quantum field theory methods.
Our results have been obtained for the continuous group $G=SU(2)$,
even  though they may be extended to more general groups. The final formulae are
given in a form where the behaviour as $k\mapsto\infty$ is obvious. In this 
connection we have explicitly exhibited the first term in the level $k$ 
asymptotic expansion for compact hyperbolic families of 3-manifolds. This 
paper has shown the validity of the asymptotic expansion for a wide class 
of hyperbolic 3-manifolds, and we hope that this analysis may be extended to 
a larger class of examples.

With regard to possible physical applications, we would like to mention that 
the evaluation of the Ray-Singer torsion presented in this paper for the
compact 3-hyperbolic manifold may be useful within the Euclidean path-integral
approach to 3-dimensional quantum gravity, where the partition function is 
evaluated by summing contributions from all possible topologies 
\cite{carl93-10-207}. In fact, for negative cosmological constant $\La$, the 
classical extrema of the Euclidean action are hyperbolic manifolds. In 
particular, we may consider a compact hyperbolic manifold. It has been 
shown that 3-dimensional gravity can be rewritten as a Chern-Simons theory for
a suitable gauge group \cite{witt88-311-46}. Therefore in the one-loop
partition function the quantum prefactor turns out to be dependent only on the 
Ray-Singer torsion of a compact hyperbolic manifold. The result of Sect. 6 
leads to the conclusion that the dependence on the volume of the Ray-Singer 
torsion is exponentially decreasing, making a contribution to the one-loop 
Euclidean partition function of the same nature of the one corresponding to 
the classical action. This fact stems from the Eq. (6.14) and the result of 
ref.\cite{carl93-10-207}, namely we have
\beq
Z_{{\cal H}^3}\equiv \tilde{\tau}^{\fr{1}{2}}({\cal H}^3)
\exp\left(-\frac{V({\cal F})}{4\pi G \sqrt |\La|}\right)\mbox{.}
\label{car}\eeq
As a consequence, the one-loop Euclidean partition function, including only 
one extremum with $\La <0$ and in absence of zero modes, reads
\beq
Z_{{\cal H}^3} \equiv \frac{({\cal Z}_1(2))^{\fr{1}{2}}}{({\cal Z}_1(1))^{\fr{1}{4}}}
\exp\left(-\frac{V({\cal F})}{4\pi G} \aq\frac{1}{G \sqrt |\La|} + \frac{1}{3}
\cq \right)\:,\label{car1}\eeq
where the second term in the exponential is the first quantum correction.
Thus, in this case the volume dependence on the partition function has been 
completely extracted and the exponential suppression, due to the large volume 
dependence, confirmed.

\ack{ The research of A.A.B.~was supported in part by Russian Foundation for
Fundamental Research grant No.~95-02-03568-a and by Russian Universities
grant No.~95-0-6.4-1.}

\appendix

\s{The Plancherel Measure Associated with Transverse Vector Fields on $H^3$}

Let $A=A_adx^a$ be a 1-form in $H^N$, while the transverse gauge is
$\nabla^aA_a=0$. Then it may always be written in the form $\nabla^a\om_{ab}$,
for some 2-form $\om_{ab}$\cite{bade89-202-431}. The eigenvalue equation for the
Laplace operator takes the form
\beq
g^{ab}\nabla_a\nabla_bA_c-R^a_cA_a=-\la A_c\mbox{.}
\label{vect}
\eeq
This operator is self-adjoint in the Hodge scalar product
\beq
(A,B)=\int_{H^N}A\wedge\ast B=\int_{H^N}g^{ab}A_aB_b\sqrt{g}d^Nx\mbox{.}
\label{scpro}
\eeq
The spectrum of the operator presented in Eq. (\ref{vect}) is continuous and
it contains a gap, since $\la\geq(\rho_N-1)^2$, where $\rho_N=(N-1)/2$. The
exception is the case $N=3$, where there are harmonic and square integrable
one-forms \cite{donn81-33-365}. Because of the gap, we use the parameter
$r^2=\la-(\rho_N-1)^2\geq0$ to label the spectrum. We define the density of
states (the Plancherel measure) $\mu_N(r)$, so that the zeta function per
unit volume \cite{will92-105,byts96-266-1}
\beq
\zeta_N(s|\lap_1)=\Tr[g_{ab}\lap-R_{ab}]^{-s}=\int_0^\ii[r^2+(\rho_N-1)^2]^{-s}
\mu_N(r)dr\mbox{,}
\label{zetavecN}
\eeq
has a residue at its most right pole as demanded by the general theory. Using
the asymptotic behaviour of the solutions, one finds the formula
\beq
\mu_N(r)=\frac{2(N-1)}{2^N\pi^{N/2}\Ga(N/2)[r^2+(\rho_N-1)^2]}\left|
\frac{\Ga(ir+\rho_N+1)}{\Ga(ir)}\right|^2\mbox{.}
\label{measurvec}
\eeq
For example, in the case of $N=2$ the measure containts a contribution of a
discrete spectrum and it can be written (perhaps improperly) as
\beq
\mu_2(r)=\frac{r}{2\pi}\tanh\pi r+\frac{1}{2\pi}\de(r-i/2)\mbox{.}
\label{mea2}
\eeq
Therefore the trace of the vector heat kernel in two dimensions is
\beq
K(t)\equiv\Tr\left(e^{-t\lap_1}\right)=\frac{1}{2\pi}\int_0^{\ii}e^
{-t(r^2+1/4)}r\tanh(\pi r)dr+\frac{1}{2\pi}\mbox{.}
\label{heat2}
\eeq

For $N=3$ the zeta function in Eq.~(\ref{zetavecN}) generally speaking do not
exist and a mass term is needed to achieve convergence. Doing this we obtain
\beq
\zeta_3(s|\lap_1)=\frac{2\varrho^{2s-3}}{(4\pi)^{3/2}\Ga(s)}\left[(m\varrho)^
{3-2s}\Ga(s-3/2)+2(m\varrho)^{1-2s}\Ga(s-1/2)\right]\mbox{,}
\label{zetavec3}
\eeq
where the curvature radius $\varrho$ has been reinserted. Note that the zero
mass limit is not uniform in $s$, i.e it depends on $s$.

For $N=4$ no mass term is needed and we get
\begin{eqnarray}
\zeta_4(s|\lap_1)&=&\frac{3\varrho^{2s-4}}{16\pi^2}\left[\frac{4^{s-2}}{(s-2)
(s-1)}+\frac{4^{s-1}}{s-1}\right] \nn \\
&-&\frac{3\varrho^{2s-4}}{4\pi^2}\int_0^\ii\left(r^2+\frac{9}{4}\right)
\left(r^2+\frac{1}{4}\right)^{-s}\frac{1}{1+e^{2\pi r}}dr\mbox{.}
\label{zetavec4}
\end{eqnarray}
It can be shown explicity that the integral in the last equation is an analytic
function of $s$.

\s{The Selberg Trace Formula and Zeta Functions}

We now specialize the results of Sect. 3 to 3-manifold with topology
${\cal H}^3$. First we remind shortly some geometric information necessary
for the trace formula.

Let $\ga$ be a closed geodesic in ${\cal H}^3$ and $l_{\ga}$ its length. Let us
consider a parallel translation of any vector along $\ga$. After the journey,
the vector will be rotated in the two-space orthogonal to the tangent vector
at $\ga$. Let $R_{\ga}$ be a corresponding $SO(2)$ matrix of rotation. Thus
every closed geodesic is associated with a certain element of $SO(2)$ and a
certain number $l_{\ga}$. If $n$ is the winding number of $\ga$ then
$R_{\ga}=R_{\de}^n$ for some rotation $R_{\de}$, and $R_{\de}$ being not a
power of any other rotation. In this case $\ga=\de^n$ and we call $\de$ a
primitive geodesic. The set of all primitive geodesics will be denoted by
${\ca P}$. For any $\ga$ we also define the factor
$S(n,l_{\ga})=\det|1-N_{\ga}^nR_{\ga}^n|$,
where $N_{\ga}=\exp l_{\ga}$. Finally, let $\ca F$ being the fundamental
domain for the group $\Ga$ (relative to the invariant Riemannian measure) and
$V(\ca F)$ is its volume. Thus one can now state the trace formula
\cite{frie86-84-523,deit89-59-101}.

\begin{proposition}
Let $h(r)$ be a function even and holomorphic in a strip
larger than $1$ about the real axis and such that $h(r)=O(r^{-3-\ep})$
uniformly in the strip as $r\rightarrow\ii$. Then

\begin{eqnarray}
\sum_jh(r_j)+(b_1-b_0)h(0)&=&V(\ca F)\int_0^\ii h(r)
\mu_3(r)dr \nn \\
&+&\sum_{\{\ga\}\in\ca P}\sum_{n=1}^{\ii}\Tr[R_{\ga}^n]
\frac{\chi^n(\ga)l_{\ga}N_{\ga}^n}
{S(n;l_{\ga})}\hat h(nl_{\ga})\mbox{,}
\label{selbvec}
\end{eqnarray}
where each number $r_j$ is the positive root of $r_j^2=\la_j$,\,\,
$\la_j$ are the eigenvalues of the Laplace operator acting on transverse
vector fields and the summation over $j$ includes all the non-vanishing
eigenvalues counted along with their degeneracy.
\end{proposition}
In addition $\hat h(p)=\frac{1}{2\pi}\int_{-\ii}^{\ii}e^{ipr}h(r)dr$ and the
Plancherel measure has the form $\mu_3(r)=(r^2+1)\pi^{-2}$.

\begin{proposition}
The Selberg trace formula for twisted scalar fields is similar and
reads \cite{selb56-20-47,byts96-266-1}
\begin{eqnarray}
\sum_jh(r_j)+b_0h(i)&=&V(\ca F)\int_0^\ii h(r)
\Phi_3(r)dr \nn \\
&+&\sum_{\{\ga\}\in\ca P}\sum_{n=1}^{\ii}
\frac{\chi^n(\ga)l_{\ga}N_{\ga}^n}
{S(n;l_{\ga})}\hat h(nl_{\ga})\mbox{,}
\label{selbsca}
\end{eqnarray}
where now each number $r_j$ is the root of $r_j^2=\la_j-1$ in the
upper half complex plane, $\la_j$ are the eigenvalues of the Laplace operator
acting on twisted scalar fields and the sum is over all the non vanishing
eigenvalues, including degeneracy.
\end{proposition}
Furthermore, the measure is $\Phi_3(r)=r^2(2\pi^2)^{-1}$.
The Selberg zeta functions we are going to introduce are important in
the following.

The $\Xi$ function for 1-forms (for 0-forms the definition is the
same, but the factor $\Tr[R_{\ga}^n]$ in the equations below is put equal to
one) can be written as
\beq
\Xi_1(s)=\sum_{\{\ga\}\in\ca P}\sum_{n=1}^{\ii}\Tr[R_{\ga}^n]
\frac{\chi^n(\ga)l_{\ga}}{S(n;l_{\ga})}\exp[-(s-2)nl_{\ga}]\mbox{,}
\label{xivec}
\eeq
while the Selberg zeta function is given by
\beq
{\cal Z}_1(s)=\exp\left\{-\sum_{\{\ga\}\in\ca P}\sum_{n=1}^{\ii}\Tr[R_{\ga}^n]
\frac{\chi^n(\ga)}{S(n;l_{\ga})n}\exp[-(s-2)nl_{\ga}]\right\}\mbox{,}
\eeq
and therefore $\Xi_1(s)={\cal Z}_1^{\prime}(s)/{\cal Z}_1(s)$. The presence
of the exponential makes it easy to expect analyticity for $\Xi_1(s)$ in the
strip $\Re s>2$. Indeed this the case, although it is far from trivial, due to
exponential proliferation of closed geodesics with increasing lengths
\cite{hube59-138-1}. It can also be shown that the analytically continued
function $\Xi_1(s)$ has simple poles at $s_j=\pm ir_j$. If there are
zero modes, then a pole also exists at $s=1$, of residue $b_1-b_0$.

\begin{remark}
The $\Xi_1(s)$ function satisfies the functional equation
\beq
\Xi_1(s+1)+\Xi_1(-s+1)=\frac{2V(\ca F)}{\pi}(s^2-1)\mbox{.}
\label{funcvec}
\eeq
The poles of $\Xi_1(s)$ become poles or zeroes of ${\cal Z}_1(s)$, depending
on the sign of the residues and the functional equation for ${\cal Z}_1(s)$
has the form
\beq
{\cal Z}_1(-s+1)={\cal Z}_1(s+1)\exp\left[\frac{2V(\ca F)}{\pi}s(1-\frac{s^2}
{3})\right]\mbox{.}
\label{zfunvec}
\eeq
\end{remark}
In particular, ${\cal Z}_1(s)$ has a pole or a zero at $s=1$ of order
$b_1-b_0$ depending on whether this number is negative or positive,
namely ${\cal Z}_1(s)=s^{(b_1-b_0)} G_1(s)$ with finite number $G_1(0)$.

Generally speaking zeta functions ${\cal Z}_p(s)$ can actually be defined
not only for 1- or 0-forms, but also for $p$-forms. In the case of 0-forms,
the defining equations are those for ${\cal Z}_1(s)$ in which the
factor $\Tr[R_{\ga}^n]$ is replaced with one.
\begin{remark}
The functional equation for the ${\cal Z}_0(s)$ is
\beq
{\cal Z}_0(-s+1)={\cal Z}_0(s+1)\exp\left[-\frac{V({\ca F})}{3\pi}s^3\right]
\mbox{.}
\label{zfuncsca}
\eeq
\end{remark}
The ${\cal Z}_0(s)$ is the entire function of order three, and all poles of
$\Xi(s)$ are zeroes of the ${\cal Z}_0(s)$, a fact which holds true in any
odd dimension. In particular, it has a zero of order $b_0$ at $s=0$,
namely ${\cal Z}_0(s)=s^{b_0} G_0(s)$ with finite number $G_0(0)$.
Finally for $p$-forms one replaces $R_{\ga}\in SO(2)$, acting on ${\bf C}^2$,
with its representation acting on the exterior algebra $\La^p{\bf C}^2$.

\end{document}